\documentclass[aps, prl, reprint, superscriptaddress, showkeys]{revtex4-2}
\pdfoutput=1

\usepackage{amsfonts,amsmath,,amssymb,
		xcolor,
		dcolumn,
		framed,
		graphicx,
		lipsum,
		moreverb, multirow,
		natbib,
		ulem,
		verbatim
		}
\usepackage{siunitx} %for the angle symbol \ang
\usepackage{bm} % bold symbols in formulas

\usepackage[pdftex,
            ]{hyperref}

\def\up{\uparrow}
\def\down{\downarrow}
\def\be{\begin{equation}}
\def\ee{\end{equation}}
\def\ber{\begin{eqnarray}}
\def\eer{\end{eqnarray}}

\def\bsigma{\mbox{\boldmath $\sigma$}}

\def\bb{{\bf b}}

\def\bk{{\bf k}}
\def\bq{{\bf q}}

\def\bn{{\bf n}}

\def\bs{{\bf s}}

\def\calG{{\mathcal G}}

\def\bbG{{\mathbb G}}

\newcommand\commentout[1]{}

\def\be{\begin{equation}}
\def\ee{\end{equation}}
\def\ber{\begin{eqnarray}}
\def\eer{\end{eqnarray}}

\def\bk{{\bf k}}

\newcommand{\ie}{{\it i.e.~}} 	%i.e.
\newcommand{\eg}{{\it e.g.~}} 	%e.g.

\definecolor{greenS}{rgb}{0.00, 0.6, 0.00}
\definecolor{orangeS}{rgb}{0.6, 0.1, 0.1}

\begin{document}

\title{Tunneling anisotropic spin galvanic effect}
\author{Genevi\`{e}ve Fleury}
\affiliation{SPEC, CEA, CNRS, Universit\'{e} Paris-Saclay, 91191 Gif-sur-Yvette, France}
\author{Michael Barth}
\affiliation{Institut f\"ur Theoretische Physik, Universit\"at Regensburg, 93040 Regensburg, Germany}
\author{Cosimo Gorini}
\email[]{cosimo.gorini@cea.fr}
\affiliation{SPEC, CEA, CNRS, Universit\'{e} Paris-Saclay, 91191 Gif-sur-Yvette, France}

\date{\today}

\begin{abstract}

We show that pure spin injection from a magnetic electrode into an inversion symmetry-broken system composed of a tunnel barrier and a metallic region generates a transverse charge current.  Such a tunneling spin galvanic conversion is robust to disorder and non-local, \ie injection and detection contacts do not coincide, and is strongly anisotropic whenever the internal spin-orbit field has a non-trivial angular dependence.  The anisotropy shows up in linear response, contrary to what happens in bulk conversion setups lacking tunnelling elements.  This is particularly relevant for spin-charge conversion at oxide interfaces, where both the tunnel barrier and the receiving low-dimensional metallic system host effective spin-orbit fields with complex angular symmetries.

\end{abstract}

\maketitle

\begin{figure}
	\includegraphics[width=\columnwidth]{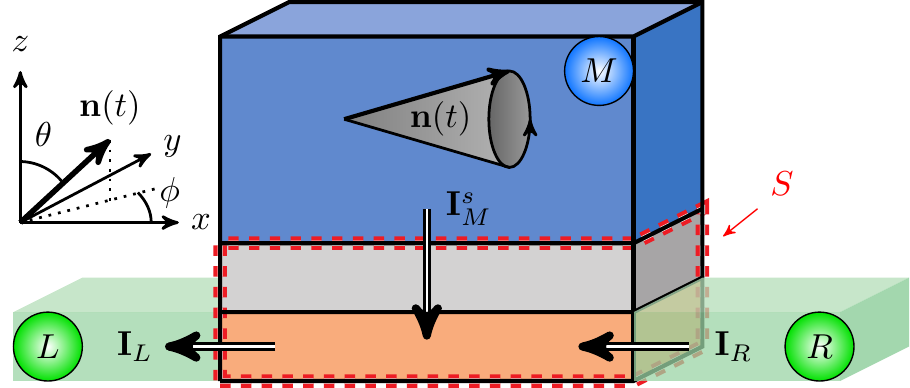}
	\caption{
	Spin pumping setup: the top magnetic $(M)$ and two left $(L)$ and right $(R)$ normal electrodes are connected to a scattering region $S$ marked by the red dashed line.  $S$ consists of a tunnel barrier (gray) on top of a metallic layer (orange).  Inversion symmetry is broken along $z$, either within the barrier, in the underlying layer or in both, yielding a SOC field \`{a} la Rashba. The $M$ electrode is driven and can inject/absorb spins but no charges (open circuit).  The spin-charge conversion voltage is measured between the $L$ and $R$ electrodes (closed circuit) as a function of $\bn_0(\theta,\phi)$, \ie the equilibrium direction of the magnetization in absence of driving.
        }
\label{fig_setup}
\end{figure}

Spin-orbit coupling (SOC) in metallic systems offers many possibilities for converting spin signals into charge ones and vice-versa \cite{zutic2004,fert2019}.  In particular, charge currents may be generated by pure spin injection via the spin galvanic effect (SGE) \cite{ganichev2002,ganichev2003,ivchenko2017,gorini2017,maleki2018} -- the conversion of a non-equilibrium spin accumulation into a charge current -- and/or the inverse spin Hall effect (ISHE) \cite{dyakonovbook,fert2019,bakun1984,valenzuela2006,hahn2013} -- the conversion of a pure spin flow into a transverse charge flow.
The working principle of the typical spin pumping setup, sketched in Fig.~\ref{fig_setup}, relies on both phenomena: a magnetic electrode is driven by microwaves, and its precessing magnetization injects angular momentum -- but on average no charges -- into an underlying metallic system, where SOC converts it into a measurable electric voltage.  Broadly speaking, there are two scenarios: (i) The receiving system is three-dimensional (3D), so that pumping results in a pure spin current flowing away from the magnet.  This is the case for popular metal-based setups, where the bulk ISHE dominates spin-charge conversion\cite{valenzuela2006,liu2012,hahn2013,obstbaum2014,karnad2018}; (ii) The receiver has no thickness through which an injected spin current may flow, \eg it is a two-dimensional electron gas (2DEG) at an interface or on the surface of a 3D topological insulator.  In this case the absorbed angular momentum builds up a spin accumulation, which is converted into a voltage by the SGE\cite{rojassanchez2013,shiomi2014,rojassanchez2016}.  If the importance of interfacial SOC \`a la Rashba \cite{bercioux2015,amin2016} is agreed upon, the situation is in practice not always that clear-cut.  This leaves room for debate concerning the dominance of specific conversion channels, as both bulk and interfacial contributions may exist and compete \cite{amin2016,hellman2017,wen2019,fert2019,pham2021}.    

%%%%%%%%%%%%%%%%%%%%%%%%%%%%%%%%%%%%%%%%%55
%%%%%%%%%%%%%%%%%%%%%%%%%%%%%%%%%%%%%%%%%

A further layer of complexity is added by the injection process itself, which happens through an inversion-asymmetric magnetic tunnel junction.  Due to the interplay of magnetism and interfacial SOC from inversion symmetry breaking, junctions of these sort host a plethora of anisotropic magneto-electric effects \cite{fabian2007,matosabiague2009,hellman2017}.  Recently, some tunneling spin Hall \cite{matosabiague2015} and anomalous Hall effects were proposed \cite{tarasenko2004,matosabiague2015}, arising from under-the-barrier transmission which is not only spin-sensitive, but also skewed in momentum space\footnote{Under-the-barrier spin-charge effects are fundamental in strongly disordered systems, see Ref.~[\onlinecite{smirnov2017}]}.  Skewedness actually appears also if SOC is present only on the injecting/receiving metallic sides \cite{dang2015,rozhansky2020,to2021}, rather than only under the barrier \cite{matosabiague2015}.  As emphasised in Ref.~[\onlinecite{to2021}], skewed injection is crucial in a novel spin-charge conversion platform rapidly on the rise: that of high-quality 2DEGs at oxide interfaces \cite{ohtomo2004}, whose fundamental and technological potential is beyond question \cite{reyren2007,vaz2019,noel2020}.  Such systems can be easily manipulated via gates and are intrinsically inversion-asymmetric, with more or less complex forms of Rashba SOC on the 2DEG side\cite{zhou2015,seibold2017,vaz2019}.  Closely related systems also host various exotic transport phenomena \cite{bousquet2008,valencia2011,noel2020}.   

A recent experiment showed that spin-charge conversion in the 2DEG at the ${\rm LaAlO}_3|{\rm SrTiO}_3$ (LAO$|$STO) interface is indeed strongly anisotropic, carrying imprints of the spin texture of the effective Rashba field \cite{elhamdi2022}\footnote{Ref.~\cite{elhamdi2022} also discusses orbital angular momentum-to-charge conversion, while here we focus on spin-charge conversion only.}.   
This contrasts with the known fact that the Onsager reciprocal phenomenon -- the generation of a non-equilibrium spin accumulation by driving a current -- is isotropic in the very same kind of systems, independently of the Rashba texture \cite{johansson2021}.  Furthermore, at an oxide interface Rashba SOC is present not only on the 2DEG side, but also in the barrier separating it from the spin pumper\cite{elhamdi2022}, and both may contribute to spin-charge (charge-spin) conversion.    
Given the context, our work addresses two central questions:

(i) How can spin-orbit coupling generate an anisotropic transverse charge current when a tunnel barrier is purely spin-biased, \ie when only angular momentum but no net charge flows through the barrier itself?  The goal is to do for the SGE what was done for
the anomalous Hall \cite{matosabiague2015,tarasenko2004} and spin Hall effects \cite{matosabiague2015}, when they were generalized to include tunneling \footnote{Ref.~\cite{tarasenko2004} calls ``tunneling spin galvanic effect'' what is later named ``tunneling anomalous Hall effect'' in Ref.~\cite{matosabiague2015}, \ie a transverse charge response due to a spin-polarized charge current through a barrier with SOC.  We find the terminology of Ref.~\cite{matosabiague2015} more appropriate, and reserve ``spin galvanic'' here to the charge response to a pure spin bias.}.  Since the three effects make up the family of the ``spin Hall effects'' \cite{engelbook}, our work closes the circle.

(ii) What is the Onsager reciprocal observable of such a tunneling spin galvanic current?

To answer these questions we build a theory framework describing spin-charge conversion in inversion symmetry-broken multi-terminal setups, in which a driven magnetic electrode acts as a pure spin injector.  Onsager reciprocity is fulfilled by construction.  The theory also treats tunnel and receiving elements on the same footing, thus including SOC and magnetism in either or both.  Motivated by a recent experiment, we apply the general theory to a model system of an oxide interface junction.  In so doing we identify a spin-charge conversion channel which mixes skew-tunneling and standard SGE physics, and which we refer to in the following as ``tunneling anisotropic SGE" -- see Eq.~\eqref{eq_sc_conductance}.  Simulations in disordered samples show that the phenomenon is robust with respect to scattering.

\begin{figure}
	\includegraphics[width=\columnwidth]{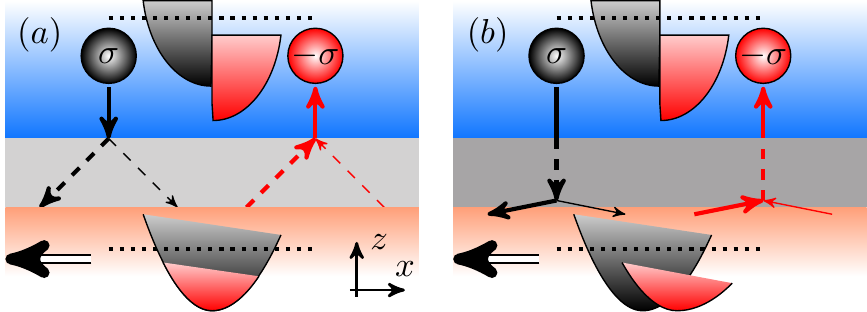}
	\caption{
	Inversion symmetry-broken magnetic tunnel junction under a pure spin bias (same color code from Fig.~\ref{fig_setup}).  The black dotted lines mark the electrochemical potential $\mu_{L,\sigma}=\mu_{L,-\sigma}=\mu_{R,\sigma}=\mu_{R,-\sigma}=\mu+\delta\mu$ of the side electrodes, which floats to ensure overall charge neutrality.}
\label{fig_mechanisms}
\end{figure}

The proposed effect should appear in any magnetic tunnel junction with broken inversion symmetry under a spin bias, since it works on the general principles sketched in Fig.~\ref{fig_mechanisms}:
Mott skew scattering results from spin-momentum correlations induced by SOC when electrons imping on impurities\cite{dyakonovbook}; Similar correlations appear if electrons cross any scattering region with SOC, \eg by tunneling through a spin-orbit-coupled barrier \cite{tarasenko2004,matosabiague2009,matosabiague2015} or by entering from/landing into a region with SOC \cite{dang2015,to2021,rozhansky2020}, leading to various spin-charge conversion channels.  (a): SOC is present only in the barrier, where a tunneling ISHE takes place: the resulting skewed populations of both spin-degenerate bands yield each a transverse charge current.  Since SOC is absent from the receiving metal, the non-equilibrium spin accumulation induced by pumping is not converted into a current, \ie there is no standard SGE on the receiving side.  (b): SOC is present only on the receiving metallic side, resulting in two SOC-split (Rashba) bands.  Skewed injection takes place at the exit of the barrier, causing an asymmetric population of both bands, each yielding a current similarly to case (a).  Such skew-tunneling-induced effect is however not all: the states are coupled by scattering according to standard SGE physics \cite{ganichev2002,ganichev2003,ivchenko2017,rousseau2021}, which now contributes to the overall SGE of the junction.  In a general inversion-asymmetric junction both (a) and (b) mechanisms are present and responsible for the ``tunneling anisotropic SGE".

%%%%%%%%%%%%%%%%%%%%%%%%%%%%%%%%%%%%%%%%%55
%%%%%%%%%%%%%%%%%%%%%%%%%%%%%%%%%%%%%%%%%

\textit{Steady state transport theory} -- 
Without loss of generality we first focus on the essentials and consider the three-terminal system from Fig.~\ref{fig_setup}.  Since we are interested in the DC output of the setup, we reformulate the time-dependent spin pumping problem as an effective steady-state problem.  This substantial simplification allows us to use time-independent scattering theory -- much simpler and numerically cheaper than any time-dependent approach.  

The magnet hosts free electrons whose spin $\bsigma$ couples to the magnetization via standard $s$-$d$ exchange $H_{xc} = -(\Delta_{xc}/2)\bn(\theta,\phi)\cdot\bsigma, |\bn|=1$.  The magnetization angles $\theta,\phi$ are defined as usual, see Fig.~\ref{fig_setup}.  Under driving the magnetization precesses, $\bn\to\bn(t)$, producing in the magnet a non-equilibrium spin polarisation (density) $\delta \bs(t)=\hbar^2 N_0/2\left[\bn\times\dot{\bn} - (\hbar/\Delta_{xc}\tau_s)\dot{\bn}\right]$, with $N_0$ the density of states per spin and unit volume at the Fermi energy, and $\tau_s$ the spin relaxation time.  Such spin polarization has a steady-state component ${\overline{\delta\bs}}=\hbar^2 N_0/2\left[\overline{\bn\times\dot{\bn}}\right]$.  The latter can be used to define an effective spin electrochemical potential proportional to the driving frequency $\omega$ \cite{toelle2017,rousseau2021}, $\hbar\,\overline{\bn\times\dot{\bn}}\equiv\delta{\boldsymbol\mu}_s.$ The spin pumper thus acts as a magnetic electrode under a pure steady-state spin bias -- a spin bias in the absence of any electric one. It is easy to see that $\delta{\boldsymbol\mu}_s$ is parallel to the (arbitrary) equilibrium direction $\bn_0$ of the magnetization, since misaligned spins relax too fast to allow any buildup \cite{toelle2017,rousseau2021}.  We thus have $\delta\mu_s=\delta\mu_M^\up-\delta\mu_M^\downarrow$, where $\delta\mu_M^\sigma, \sigma=\up,\down$ is the deviation from equilibrium of the electrochemical potential for $\bn_0$-polarized majority/minority electrons. The $\sigma$-polarized currents flowing into/out of the 3-terminal setup of Fig.~\ref{fig_setup} are written in Landauer-B\"uttiker form following Ref.~[\onlinecite{kara2020}]
\be
I^\sigma_\alpha = \frac{e}{h}\sum_{\beta,\sigma'=\up\down} \int\,{\rm d}\epsilon%\inten 
\left[
 f(\epsilon,\mu_\alpha^\sigma) - f(\epsilon,\mu_\beta^{\sigma'})
\right]
T^{\sigma\sigma'}_{\alpha\beta}(\epsilon).
\ee
Here $T_{\alpha\beta}^{\sigma\sigma'}$ is the transmission probability from lead $\beta$ with spin $\sigma'$ to lead $\alpha$ with spin $\sigma$, and $f(\epsilon,\mu)=1/\left[1+e^{(\epsilon-\mu)/(k_BT)}\right]$ is the Fermi function, $T$ being the (uniform) temperature and $k_B$ the Boltzmann constant.  In our configuration the left and right ($\alpha=L,R$) normal electrodes are at the same electrochemical potential $\mu_{L}^{\sigma} =\mu_{L}^{\bar{\sigma}}=\mu_{R}^{\sigma}=\mu_{R}^{\bar{\sigma}}\equiv\mu+\delta\mu$, with $\bar{\sigma}\equiv-\sigma$, while in the magnetic ($\alpha=M$) terminal $\mu_{M}^{\sigma}=\mu+\delta\mu_M^\sigma$.
Linear response (small $\delta\mu$, $\delta\mu_M^\sigma$) yields in the normal electrode
$I^\sigma_\alpha = \sum_{\sigma'}[(\delta\mu_M^{\sigma'}-\delta\mu)/e] G_{\alpha M}^{\sigma\sigma'},\,\alpha = L, R$,
%\be
%I^\sigma_\alpha = \sum_{\sigma'}
%\left(
%\frac{\delta\mu_M^{\sigma'}-\delta\mu}{e}
%\right)
%G_{\alpha M}^{\sigma\sigma'},
%\quad
%\alpha = L, R
%\label{eq_INR_LinRes}
%\ee
%in the normal electrodes and in the magnetic one
and in the magnetic one $ I^\sigma_M = [(\delta\mu-\delta\mu_M^{\sigma})/e]\sum_{\sigma'}[G_{ML}^{\sigma\sigma'}+G_{MR}^{\sigma\sigma'}]+ [(\delta\mu_M^{\bar{\sigma}}-\delta\mu_M^{\sigma})/e]G_{MM}^{\sigma\bar{\sigma}}$.
%\begin{align}
%I^\sigma_M = &\left(\frac{\delta\mu-\delta\mu_M^{\sigma}}{e}\right)\sum_{\sigma'}[G_{ML}^{\sigma\sigma'}+G_{MR}^{\sigma\sigma'}] \nonumber\\
%&+ \left(\frac{\delta\mu_M^{\bar{\sigma}}-\delta\mu_M^{\sigma}}{e}\right)G_{MM}^{\sigma\bar{\sigma}}
%\label{eq_INT_1}
%\end{align}
%
The conductances are $G_{\alpha\beta}^{\sigma\sigma'} = \frac{e^2}{h}\int\,{\rm d}\epsilon(-\partial_\epsilon f_0) T_{\alpha\beta}^{\sigma\sigma'}\,$, with $f_0$ the equilibrium distribution.
%\be
%\label{eq_conductances}
%G_{\alpha\beta}^{\sigma\sigma'} = \frac{e^2}{h}\int\,{\rm d}\epsilon\left(-\frac{\partial f_0}{\partial \epsilon}\right) T_{\alpha\beta}^{\sigma\sigma'}\,.
%\ee

Charge conservation dictates that the currents $I_\alpha = \sum_\sigma I_\alpha^\sigma$ add up to zero, $I_M + I_L + I_R = 0$.  Furthermore, the pumping electrode remains charge neutral on average, \ie $I_M=0$ \footnote{See Ref.~[\onlinecite{rouzegar2021}] for a similar discussion in a related context.}.  Lengthy but straightforward calculations allow to write the spin current in lead $M$ $(I^s_M\equiv (\hbar/2e)\left[I^\up_M - I^\down_M\right])$ and the charge currents in the normal leads $(I_L, I_R)$ in response to the spin bias $\delta\mu_s$.  The spin-charge conversion current $I_{sc}$ generated by the tunnelling anisotropic SGE is a transverse current.  We define it as the difference between L and R currents $I_{sc} = I_L - I_R = \calG_{sc}\delta\mu_s/e.$
%\be
%\label{eq_sc1}
%I_{sc} = I_L - I_R = \calG_{sc}\delta\mu_s/e.
%\ee 
The corresponding conductance $\calG_{sc}$ reads
\be
\label{eq_sc_conductance}
\calG_{sc} =
\frac{
\bbG_{LM}\left(\bbG^\uparrow_{RM}-\bbG^\downarrow_{RM}\right) - \bbG_{RM}\left(\bbG^\uparrow_{LM}-\bbG^\downarrow_{LM}\right)
}
{
\bbG_M
},
\ee
having defined $\bbG_{\alpha M}^\sigma=\sum_{\sigma'} G_{\alpha M}^{\sigma'\sigma}$, $\bbG_{\alpha M}=\sum_\sigma\bbG_{\alpha M}^\sigma$, $\bbG_M = \sum_\alpha \bbG_{\alpha M}$, with $\alpha = R, L$.
 
In the Onsager reciprocal scenario an electric bias drives a current $R\to L$, $\mu_L-\mu_R = -\delta\mu_{LR}$ \footnote{We choose the energy reference so that $\mu_L = \mu_0 -\delta\mu_{LR}/2, \mu_R=\mu_0 + \delta\mu_{LR}/2$.}, which generates a pure spin current $I^s_M$ into the M electrode.  The latter is $I^s_M = (\hbar/2e)I_{cs}$, with $I_{cs}$ the charge-spin $(cs)$ conversion current $I_{cs} = I^\uparrow_M - I^\downarrow_M = \calG_{cs}\,\delta\mu_{LR}/e.$
%\be
%I_{cs} = I^\uparrow_M - I^\downarrow_M = \calG_{cs}\,\delta\mu_{LR}/e.
%\ee
The conductance is
\be
\label{eq_cs_conductance}
\calG_{cs} = -
\frac{
\bbG_{ML}\left(\bbG^\uparrow_{MR}-\bbG^\downarrow_{MR}\right) - \bbG_{MR}\left(\bbG^\uparrow_{ML}-\bbG^\downarrow_{ML}\right)
}
{
\bbG_M
},
\ee
with $\bbG^\sigma_{M\alpha}=\sum_{\sigma'}G^{\sigma\sigma'}_{M\alpha}, \alpha=L,R$. From microreversibility in the presence of exchange interaction one has $G^{\sigma\sigma'}_{\alpha\beta}(\Delta_{xc}) = G^{\bar{\sigma'}\bar{\sigma}}_{\beta\alpha}(-\Delta_{xc})$ \cite{zhai2005,jacquod2012} which leads to the Onsager-Casimir relation for the tunneling SGE
\be
\label{eq_Onsager}
\calG_{sc}(\Delta_{xc})=\calG_{cs}(-\Delta_{xc}).
\ee
Eqs.~\eqref{eq_sc_conductance}, \eqref{eq_cs_conductance} and \eqref{eq_Onsager} are central results of our work.  They are fully general, \ie independent of any detail of the multi-terminal structure, and their extension to an arbitrary number of electrodes is straightforward.  Indeed, we verify Eq.~\eqref{eq_Onsager} for our LAO$|$STO model in a 5-terminal configuration below.

\begin{figure}
	\includegraphics[width=\linewidth]{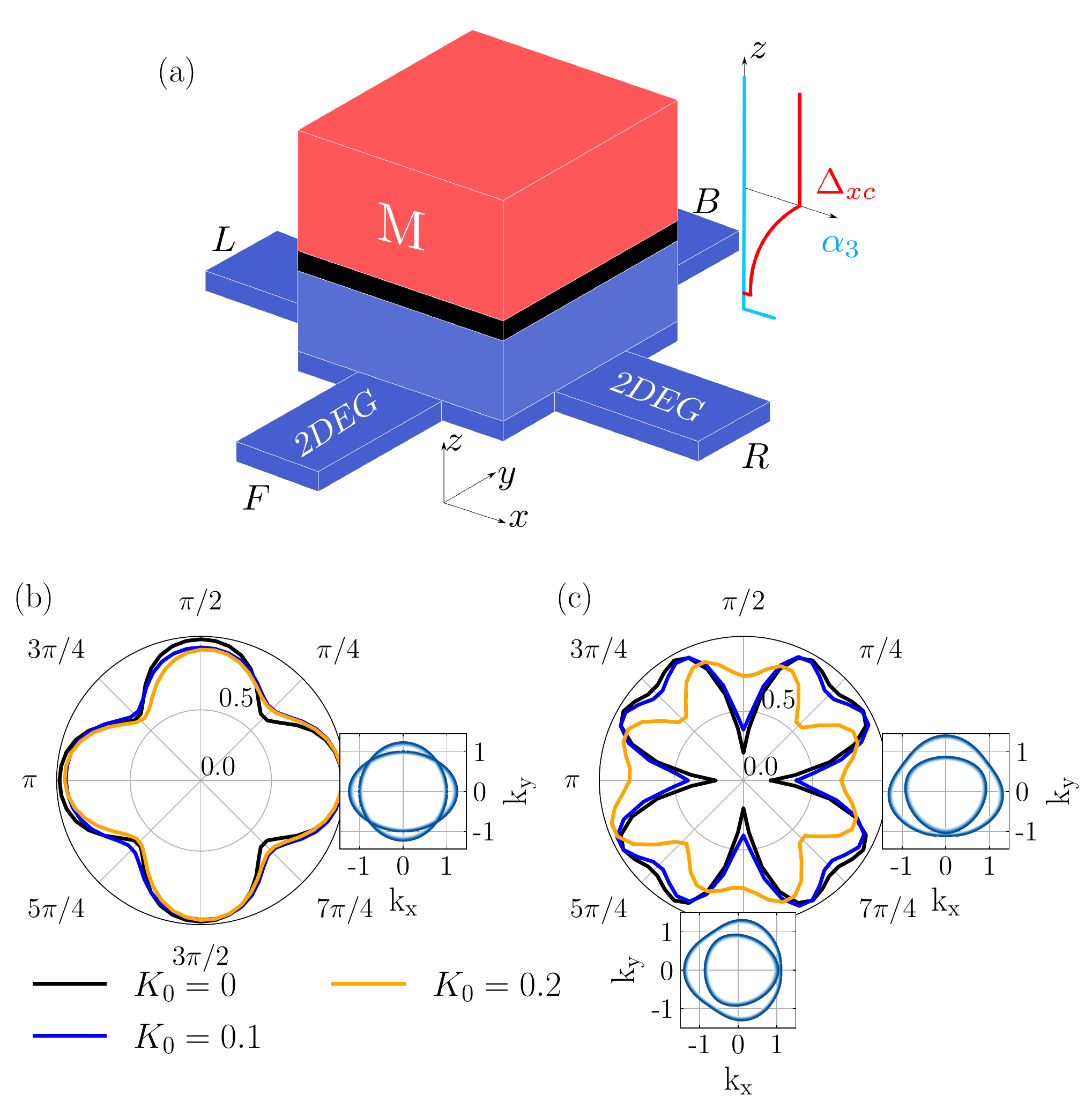}
	\caption{(a) 5 terminal setup, with sketch of the $\Delta_{xc}(z), \alpha_3(z)$ profiles. (b)-(c) Polar plots of the normalized spin-charge conversion conductance $g_{sc}(\phi) \equiv \sqrt{(\calG_{sc}^x)^2+(\calG_{sc}^y)^2}/\sqrt{(\calG_{sc}^x)_{\rm max}^2+(\calG_{sc}^y)_{\rm max}^2}$ for $\theta=\pi/2$. Panels (b), (c) show numerical data respectively for $\xi_{xc}=1.2L_z, \,1.6L_z$. Other parameters are specified in the main text. Insets: Fermi contour for $\Delta_{xc}(0)=0$ in (b) and $\Delta_{xc}(0)=-0.6$ in (c) at angles $\phi=0$ (right) and $\phi=3\pi/2$ (bottom).}
\label{fig_numerics}
\end{figure}

\textit{Anisotropies} -- To explain anisotropic effects in two-terminal magnetic tunnel junctions with SOC, Refs.~[\onlinecite{matosabiague2009,matosabiague2015}] give arguments which can be generalized to multi-terminal setups, Figs.~\ref{fig_setup},~\ref{fig_numerics}.  To be definite consider the spin-resolved transmission $\bbG_{\alpha M}^\sigma=\sum_{\sigma'}G^{\sigma'\sigma}_{\alpha M}$, written as 
\be
\bbG_{\alpha M}^\sigma = \frac{e^2}{h}
\int\,{\rm d}\epsilon \left(-\frac{\partial f_0}{\partial\epsilon}\right)
\sum_\bk W^\sigma_\alpha(\epsilon,\bk).
\ee
Here $\bk$ labels the propagating modes in lead $M$, \ie $\bk$ is momentum in the x-y junction plane. To establish direct contact with Refs.~[\onlinecite{matosabiague2009,matosabiague2015}] we introduced the spin- and momentum-resolved transmission probability $W^\sigma_\alpha(\epsilon,\bk)=\sum_{\sigma'}\sum_{\bq_\alpha} [t^\dagger t]^{\sigma\sigma'}_{\bk\bq_\alpha}$, with ${\bq_\alpha}$ the mode label (transverse momentum) in lead $\alpha$, and $t$ the transmission amplitudes entering the scattering matrix \cite{jacquod2012,kara2020}.  Without SOC the transmission $W^\sigma_\alpha(\epsilon,\bk)$ is even in $\bk$, $W^\sigma_\alpha(\bk)=W^\sigma_\alpha(-\bk)$.  With SOC in the scattering region $S$ -- either in the barrier, in the 2DEG, or in both -- there appears a SOC field $\bb(\bk)$ such that $\bb(\bk)=-\bb(-\bk)$, spoiling the $\bk\to-\bk$ symmetry of $W^\sigma_\alpha(\epsilon,\bk)$: transmission is now in general skewed.  Indeed, $W^\sigma_\alpha(\epsilon,\bk)$ is a function of the angle between $\bn$ and $\bb(\bk)$, the magnetization and SOC field defining the two physically preferred directions of the problem.  Simple manipulations show that such properties are transferred to the conductance $\calG_{sc}$, yielding the formal expansion 
\be
\calG_{sc} = \sum_\bk\sum_n\,\calG_{sc}^{(n)}\,\left[\bn\cdot\bb(\bk)\right]^n.
\ee
Odd terms vanish, while the surviving even ones reflect the spin texture defined by $\bb(\bk)$.  That is, spin-charge conversion by (tunnel) injection through $S$ is anisotropic, and the anisotropy is dictated by the shape of $\bb(\bk)$. Note that if magnetism extends into the SOC region it will modify $\bb(\bk)$ and thus the anisotropy, as shown below.
These arguments are general but qualitative, as the coefficients of the expansion are unknown.  For more quantitative statements we turn to microscopic simulations. 

\textit{Numerics: LAO$|$STO junction} -- We consider the 5-terminal configuration of a recent experiment \cite{elhamdi2022}, see Fig.~\ref{fig_numerics}~(a): the bottom 2DEG ($z=0$) is in contact with the upper magnetic electrode ($z > L_z$) via an extended barrier ($0<z\leq L_z$). Given the existing effective models for LAO$|$STO 2DEGs \cite{zhou2015,seibold2017,vaz2019}, we focus on the $d_{xz}$-$d_{yz}$ hybrid band to highlight the anisotropic character of tunneling spin galvanic physics in a minimal 2-band model.  The effective Hamiltonian includes a 4-fold symmetric cubic Rashba term \cite{zhou2015} and reads 
\begin{align}
H = &\left[\frac{p^2}{2m} + U(z)\right] + \alpha_3(z) \left(p_x^2-p_y^2\right) \left(\sigma^x p_y-\sigma^y p_x\right) \nonumber\\
&-\frac{\Delta_{xc}(z)}{2}\bn(\theta,\phi)\cdot\bsigma -\frac{\hbar^2\partial_z^2}{2m}. \label{eq_H_eff}
\end{align}
The Rashba constant $\alpha_3(z)\neq 0$ in the 2DEG and vanishes for $z>0$. The $s$-$d$ exchange term $\Delta_{xc}(z)$ is instead at full strength in the magnetic electrode, $\Delta_{xc}(z> L_z)=\Delta_{xc}$, and drops to zero towards the 2DEG, $\Delta_{xc}(0\leq z\leq L_z)=\Delta_{xc}\exp[-(L_z-z)/\xi_{xc}]$. The tunnel barrier $U(z)$ is a rectangular barrier of height $U_0$, shown in black in Fig.~\ref{fig_numerics}~(a). The 3-dimensional scattering region is built by discretizing the Hamiltonian \eqref{eq_H_eff} on a cubic lattice of size $L_x=L_y=50$ sites and height $L_z=6$ sites.  The $z=0$ layer -- the 2DEG --  is connected to four 2-dimensional leads along $x$ and $y$, all normal $(\Delta_{xc}=0, \alpha_3=0)$.  Each lead is $W_L=30$ sites wide and attached centrally to the 2DEG layer.  The upper contact is the $M$ electrode ($\Delta_{xc}\neq0, \alpha_3=0$).  In a real setup the 2DEG modes have a finite extension along $z$, which allows coupling through the barrier and into M.  To mimic this extension we model the extended barrier defined above as 3 transition layers (without SOC) just above the 2DEG, topped with two layers with on-site energy $U(z)=U_0>\mu$ representing the tunnel barrier \footnote{See Ref.~\cite{to2021} for a more realistic but numerically costlier tunnel barrier model.}. With lattice spacing $a=1$, we set the isotropic hopping parameter $t=1$, and fix $\mu=1.1, \alpha_3=-0.2, \Delta_{xc}=-0.6, U_0=1.9, \xi_\alpha=2L_z$.  For $z=0$ the on-site energy is $4t$, ensuring good coupling to the 2-dimensional leads, while it is $6t$ in the upper layers.

We use the KWANT package\,\cite{groth2014} to compute the transmissions $T^{\sigma\sigma'}_{\alpha\beta}$ at energy $\mu$.  Details are found in the Supp.~Mat.~\cite{SM}. The lead indices $\alpha, \beta$ are shown in Fig.~\ref{fig_numerics}, with $B, F$ respectively labelling the $B$(ack) and $F$(ront) contacts.  The resulting spin-charge conductance $\mathcal{G}^x_{sc}$ along the $x$-axis,
\begin{align}
\label{eq_sc_cond_5terms}
  \calG_{sc}^x =
\frac{1}{\bbG_M} & 
\left[\bbG_{LM}\left(\bbG^\uparrow_{RM}-\bbG^\downarrow_{RM}\right) - \bbG_{RM}\left(\bbG^\uparrow_{LM}-\bbG^\downarrow_{LM}\right) \right. \nonumber \\
& + \left(\bbG^\downarrow_{LM}-\bbG^\downarrow_{RM}\right) \left(\bbG^\uparrow_{BM}+\bbG^\uparrow_{FM}\right)  \nonumber \\
& - \left. \left(\bbG^\uparrow_{LM}-\bbG^\uparrow_{RM}\right) \left(\bbG^\downarrow_{BM}+\bbG^\downarrow_{FM}\right) \right],
\end{align}
is calculated at zero temperature and yields the current $I_{sc}^x = I_L - I_R = \calG^x_{sc}\delta\mu_s/e$. The $y$-current $I_{sc}^y = I_B - I_F = \calG^y_{sc}\delta\mu_s/e$ follows by exchanging $L\leftrightarrow B$, $R\leftrightarrow F$ in Eq.\eqref{eq_sc_cond_5terms}.
Results for $\xi_{xc}=1.2L_z$, which ensures that magnetism is absent from the 2DEG, are shown in Fig.~\ref{fig_numerics}~(b) (black line) as a polar plot. The spin-charge conductance shows the 4-fold symmetry of the Rashba bands ($C_{4v}$), see inset. This is compatible with experimental observations \cite{elhamdi2022}. Quantitative comparisons should however be avoided, since they require a multi-band model and orbital effects beyond our scope.  Furthermore, if magnetic exchange below the barrier grows stronger the response is distorted, see Fig.~\ref{fig_numerics}~(c), where $\xi_{xc}=1.6L_z$. The competition between SOC and magnetism splits the spin-charge conversion maxima, reflecting the distorted Fermi contours shown in the insets of panel (c).
%%%%%%%%%%%%%%%%%%%%%%%%%%%%%%%%%%%%%%%%%55
%%%%%%%%%%%%%%%%%%%%%%%%%%%%%%%%%%%%%%%%%

Figures~\ref{fig_numerics} (b) and (c) show that our results are robust to scattering.  We average over 150 configurations of standard white noise disorder $V(x,y,z)=K_0 t V_{xyz}$, with $K_0$ a dimensionless parameter setting its strength as a fraction of the hopping parameter $t$, and $V_{xyz}$ a normally-distributed random number centered on $0$.  The petal-shaped curves are perfectly visible.  As expected, convergence is better for weaker disorder (compare blue and yellow curves) and weaker magnetization (compare left and right panels).  Note that we show normalized curves, since a reliable estimation of the disorder-dependent amplitude of the bulk signal requires a more precise barrier model and a larger sample.

\textit{Onsager reciprocity and anisotropies} -- The above clarifies why in the very same oxide 2DEG the ISGE is isotropic \cite{johansson2021}, while the SGE measured in a spin pumping setup is not: Onsager reciprocal quantities are not the current-induced spin polarisation on the 2DEG side, $\delta{\bf S}_{\rm 2DEG}$, and the spin polarisation-induced current on the same side, $I_{\rm 2DEG}$.  They are rather the current $I_{\rm 2DEG}$ and the non-equilibrium spin polarisation $\delta{\bf S}_M$ on the magnetic electrode side\footnote{The polarizations are integrals over the 2DEG surface/M electrode volume of the respective surface/volume densities $\delta\bs_{\rm 2DEG}, \delta\bs_{M}$.}, \ie the whole experimental setup should be considered when discussing spin-charge reciprocity\cite{wang2012,gorini2012}. The Landauer-B\"uttiker approach does this by default \footnote{Reciprocity in the presence of SOC is treatable with other approaches, see \eg \cite{sugimoto2006,gorini2012,shen2014,smirnov2017,gorini2017} and references therein.  There is also a proposal for some form of equilibrium spin galvanic conversion, see \cite{wang2010}.  For more examples of non-equilibrium Landauer-B{\"{u}}ttiker calculations instead see \cite{adagideli2007,adagideli2012}.}.

\textit{Conclusions} -- We identified and microscopically characterised the tunnelling anisotropic SGE taking place at an inversion symmetry-broken magnetic tunnel junction under a pure spin bias, as well as its reciprocal effect.  Our theory is general and should be relevant in any multi-terminal junction where magnetization and spin-orbit coupling coexist. When applied to the specific case of an oxide-based spin pumping setup it provides a microscopic description of anisotropies of the kind recently observed and validates general phenomenological arguments \cite{elhamdi2022}. We expect our framework and conclusions to apply to orbital angular momentum-to-charge conversion as well, which can be tackled following \eg Ref.~\cite{shi2007}.  However a precise test for LAO$|$STO first requires a fully established low-energy model (see Ref.~\cite{trama2022} for a recent attempt at building one). 

\textit{Acknowledgements} -- CG thanks Michel Viret and Alexander Smogunov for helpful discussions, and the STherQO members for useful comments.  MB acknowledges support by the Deutsche Forschungsgemeinschaft (DFG, German Research Foundation) within Project-ID 314695032 -- SFB 1277 (project A07).

%%%%%%%%%%%%%%%%%%%%%%%%%%%%%%%%%%%%%%%%%%%%%%%%%%%%%%%%%%%%%%
%%%%%%%%%%%%%%%%%%    APPENDIX   %%%%%%%%%%%%%%%%%%%%%%%%%%%%%
%%%%%%%%%%%%%%%%%%%%%%%%%%%%%%%%%%%%%%%%%%%%%%%%%%%%%%%%%%%%%%

\appendix
\section{Appendix: Tight-binding model}

The five-terminal system sketched in Fig.~\ref{fig_numerics}(a) is modeled as follows. Space is discretized on a cubic grid with lattice spacing $a=1$. Each site ${\bf{r}}_i$ of coordinates $(x_i=ia,y_j=ja,z_k=ka)$ is labeled alternatively by the triplet $(i,j,k)$. The cubic scattering region $S$ is made of a 2DEG portion of length $L_x$ and width $L_y$ lying in the plane $z=0$, topped with $L_z-1$ layers. Its tight-binding Hamiltonian reads 
\begin{align}
    H_S = &   \sum_{{\bf{r}}\in S} c_{{\bf{r}}}^\dagger [\epsilon_0(z)+U(z)-\frac{\Delta_{xc}(z)}{2}{\bf{n}}(\theta,\phi)\cdot{\bsigma}] c_{{\bf{r}}} \nonumber\\
    & -t\sum_{\langle {\bf{r}}, {\bf{r}}' \rangle} c_{{\bf{r}}'}^\dagger  c_{{\bf{r}}}  + H_{SOC} 
\end{align}
where $t$ is the nearest neighbor hopping term, $\epsilon_0(z=0)=4t$ while $\epsilon_0(z\neq 0)=6t$, $U(z)=U_0$ in the two top layers representing the tunnel barrier and 0 elsewhere, and $\Delta_{xc}(z)=\Delta_{xc}\exp[-(L_z-z)/\xi_{xc}]$ is the exchange term. The cubic Rashba SO coupling is accounted for by
\begin{align}
H_{SOC} = -i&\sum_{i,j,k} \frac{\alpha_3(z_i)}{8}\,
  \left[c^\dagger_{i+3,j,k}\sigma_y + c^\dagger_{i,j+3,k} \sigma_x \right . \nonumber\\
    &-4\, c^\dagger_{i+1,j+1,k} (\sigma_x+\sigma_y)+4\, c^\dagger_{i+1,j-1,k} (\sigma_x-\sigma_y) \nonumber\\
    & \left .+\, 5\,c^\dagger_{i+1,j,k} \sigma_y + 5\,c^\dagger_{i,j+1,k} \sigma_x
 \right] c_{i,j,k}+ h.c.
\end{align}
obtained by discretizing the second term on the right-hand side of Eq.~\eqref{eq_H_eff} and we take $\alpha_3(z)=\alpha_3$ if $z=0$ and $\alpha_3(z)=0$ otherwise. Note that fine features of the polar profile of the spin-charge conductance obtained from this model are sensitive to the spatial extension and strength of $\Delta_{\rm xc}(z)$.  In the equations above, $c_{{\bf{r}}}\equiv c_{i,j,k}\equiv(c_{{\bf{r}}}^\uparrow,c_{{\bf{r}}}^\downarrow)$, $c_{{\bf{r}}}^\sigma$ being the annihilation operator of an electron at site ${\bf{r}}$ with spin $\sigma$ with respect to the $z$ direction. The sum $\sum_{\langle {\bf{r}}, {\bf{r}}' \rangle}$ is restricted to nearest neighbors.

The scattering region $S$ is then attached with nearest neighbor hopping term $t$ to four left ($L$), right ($R$), back ($B$), and front ($F$) leads of width $W_L$ in the plane $z=0$ (through the 2DEG) and to a fifth ferromagnetic lead $M$ along $z>0$ (through the barrier) as shown in Fig.~\ref{fig_numerics}(a). Their tight-binding Hamiltonians read
\begin{align}
       H_\alpha = &-t\sum_{\langle {\bf{r}}, {\bf{r}}\,' \rangle} c_{{\bf{r}}'}^\dagger  c_{{\bf{r}}}  + 4t\sum_{{\bf{r}}\in\alpha} c_{{\bf{r}}}^\dagger  c_{{\bf{r}}}
\end{align}
for $\alpha=L$, $R$, $B$, $F$, and
\begin{align}
    H_M= & -t\sum_{\langle {\bf{r}}, {\bf{r}}' \rangle} c_{{\bf{r}}'}^\dagger  c_{{\bf{r}}}  + \sum_{{\bf{r}}\in M} c_{{\bf{r}}}^\dagger [6t-\frac{\Delta_{xc}}{2}{\bf{n}}(\theta,\phi)\cdot{\bsigma}] c_{{\bf{r}}}
\end{align}
for the ferromagnetic lead. Note there is no cubic SOC in the leads and the exchange term amplitude is constant in $M$.

Let us conclude by emphasizing that the transition layers in the $z$ direction between the 2DEG and the FM electrode are there for numerical purposes: they mimick the vertical extension in real space of the 2DEG states, ensuring their good coupling with the top FM modes.  In principle one could consider a $z$-dependent Rashba constant $\alpha(z)$ smoothly decaying to zero towards the FM to further improve the matching, but this is not needed in our case.  Nevertheless, if opting for such a solution one should keep in mind the purely numerical origin of the layers.  That is, the physical system to simulate is not one composed of superimposed Rashba 2DEGs with different $\alpha$ constants. As a side remark, a more realistic (and numerically heavier) model of the tunnel junction avoiding the transition layers altogether is \eg the multi-band one from Ref.~\cite{to2021}, where $\alpha$ does not even appear; an even more accurate (and even heavier) one would start from \textit{ab-initio} calculations of the LAO$|$STO orbitals.  A quantitative treatment of the barrier is however not the subject here.

%%%%%%%%%%%%%%%%%%%%%%%%%%%%%%%%%%%%%%%%%%%%%%%%%%%%%%%%%%%%%%%%%%%%%%%%%%%%%%%%%%%%%%
%%%		Bibliography
%%%%%%%%%%%%%%%%%%%%%%%%%%%%%%%%%%%%%%%%%%%%%%%%%%%%%%%%%%%%%%%%%%%%%%%%%%%%%%%%%%%%%%

\bibliographystyle{apsrev4-2}
\bibliography{SGE_biblio_PRL}

\end{document}